\documentclass[a4paper,usenatbib]{mnras}

\usepackage{newtxtext,newtxmath}
\usepackage{amsmath}

\usepackage[T1]{fontenc}
\usepackage{ae,aecompl}
\usepackage[dvipsnames]{xcolor}

\usepackage{graphicx}	
\graphicspath{{./img/}}

\usepackage{amsmath}	

\usepackage{gensymb}

\newcommand{\bs}{\boldsymbol}

\title[Planetesimal mass distribution variability]{Streaming instability on different scales. I. Planetesimal mass distribution variability}

\author[J. J. Rucska and J. W. Wadsley]{
J. J. Rucska,$^{1}$\thanks{E-mail: rucskajj@mcmaster.ca}
J. W. Wadsley$^{1}$
\\
$^{1}$Department of Physics and Astronomy, McMaster University, Hamilton, L8S 4M1, Canada
}

\date{Accepted XXX. Received YYY; in original form ZZZ}

\pubyear{2020}

\begin{document}
\label{firstpage}
\pagerange{\pageref{firstpage}--\pageref{lastpage}}
\maketitle

\begin{abstract}
We present numerical simulations of dust clumping and planetesimal formation initiated by the streaming instability with self-gravity. We examine the variability in the planetesimal formation process by employing simulation domains with large radial and azimuthal extents and a novel approach of re-running otherwise identical simulations with different random initializations of the dust density field. We find that the planetesimal mass distribution and the total mass of dust that is converted to planetesimals can vary substantially between individual small simulations and within the domains of larger simulations. Our results show that the non-linear nature of the developed streaming instability introduces substantial variability in the planetesimal formation process that has not been previously considered and suggests larger scale dynamics may affect the process.
\end{abstract}




\section{Introduction}

Planet formation requires solid growth over a dozen orders of magnitude, from micron-sized grains embedded in protostellar clouds to centimetre or ten-centimetre sized dust pebbles in protoplanetary disks to terrestrial planets and planetary cores thousands of kilometres across. It is widely accepted that the first stage of growth, from micron-sized grains to centimetre-sized pebbles, is achieved by collisions. Similarly, once a large population of kilometre and tens of kilometre-sized planetesimals are present, these objects will interact gravitationally to build protoplanets and the final planetary system \citep{Armitage}. The intermediate growth phase, from centimetre sized pebbles to kilometre sized planetesimals, however, faces two key constraints known as the metre-barrier. 

The first barrier is rapid radial drift. All solid material feels a headwind as it orbits through the gaseous component of the disk. The gas orbits at sub-Keplerian speeds due to a radial pressure gradient, while dust attempts to orbit at the Keplerian speed. This headwind removes angular momentum from the dust, so that the dust orbit decays towards the star with a net inward radial drift. This effect is small for micron-sized dust grains that are tightly coupled to the gas, as well as for kilometre-sized objects. However, for intermediate sized objects, near one-metre, the radial drift timescale can be as short as a few hundred years \citep{Weiden77}.

The second barrier is related to collisional growth. Relative velocities in collisions between dust grains are strongly dependent on their size.  When the objects approach one metre in size, the combination of turbulence and lower drag leads to fast collisions that are always destructive, resulting in net mass loss for both objects \citep{Zsom10,Windmark12}.

These barriers act to exclude metre-sized objects from the disk. The formation of kilometre-sized planetesimals thus requires a specific mechanism that is capable of rapidly concentrating solid mass without relying on collisions between dust grains. 

\subsection{The Streaming Instability and Planetesimal Formation}

The streaming instability (SI) \citep{YG05} provides a promising mechanism to enhance dust concentrations. The SI is always present in shearing, dust-gas mixtures. It is one of a class of resonant drag instabilities (RDI) present in protoplanetary disks \citep{SquireHop18,SquireHop20}. At high dust to gas ratios it can operate faster than radial drift timescales \citep{YG05,YJ07}. 

The formation of planetesimals via the SI requires local dust densities that exceed the Roche density \citep{Li19}, so that they can condense under their own gravity.  Localized collapse occurs at local dust surface densities 2-3 orders of magnitude larger than the local average in the disk.  This represents a non-linear, evolved state of the SI that must be treated numerically \citep{YJ07,Bai101}.   Prior work has established that the non-linear phase consistently produces azimuthally oriented (i.e. globally ring-like) dust filaments  \citep{Johansenetal07,Bai102,Yang14,Simon16,Simon17,Li18}.  

In an influential paper, \citet{Johansenetal07} showed that these filaments can produce local dust densities high enough to initiate gravitational collapse and planetesimal formation. The timescale for this process is just tens of orbits.  This result highlighted the promise of SI for overcoming the metre barrier.  3D hydrodynamical simulations of shearing patches of protoplanetary disks are now well-established as a way to predict the properties of planetesimals formed by the non-linear SI \citep{Johansen09, Johansen12, Johansen15, Simon16, Simon17, Schafer17, Abod19, Li19, Nesvorny19, Gole20}.  These studies have explored how this process depends on parameters such as the dust mass \citep{Johansen09, Simon17}, dust grain size \citep{Simon17}, radial pressure gradient \citep{Abod19} and local gas turbulence \citep{Gole20}.

Ideally, the streaming instability would operate directly within simple (e.g. smooth, axisymmetric) models based on observations of protoplanetary disks.  However, achieving growth rates relevant to planetesimal formation may require local dust-to-gas mass density ratios greater than unity \citep{YG05,YJ07}. In simulations of local patches of protoplanetary disks this translates to a requirement of super-solar dust-to-gas surface densities in order to achieve sufficient dust clumping for gravitational collapse \citep{JY09,Bai102, Bai103}. Local concentrations of dust in the disk would circumvent this issue.  Large-scale gas structures such as pressure bumps and vortices could create large scale dust traps with enhanced local dust-to-gas mass surface density ratios \citep[see][for a review]{Birnstieletal16}. Observations show protoplanetary disks in nature can have non-uniform dust distributions, including rings \citep[e.g.][]{DSHARP18} and non-axisymmetric bumps \citep{vdM13,vdM15}\footnote{Note: features in the dust surface density formed directly by the non-linear SI are much too small to be observed directly.}. \citet{Draz14} and \citet{Draz16} presented global models of dust in protoplanetary disks using semi-analytic prescriptions for planetesimal formation via the SI, and conclude that planetesimal formation via the SI is most efficient in regions with enhanced solid abundances such as beyond the snow line, or where dust pebbles can accumulate due to radial drift pile-up.

Planetesimals formed by the streaming instability are sand-piles and initially lack cohesion other than their own self-gravity.  This fits the emerging consensus that asteroids are rubble piles and represent somewhat evolved planetesimals \citep{Walsh18}. For example, data from the recent fly-by of the New Horizon's space mission of Kuiper Belt object 486958, Arrokoth, supports the gravitational collapse scenario. \citet{McKinnon20} and \citet{Grishin20} report that this object, which is characterized by two distinct lobes, was likely formed by a low-velocity impact resulting from the slow decay of a binary orbit of two smaller Kuiper Belt objects. Additionally, \citet{Nesvorny19} compared the observed distribution of prograde vs. retrograde binary orbits in trans-Neptunian objects with similar, planetesimal-sized objects formed via the SI in local simulations of patches of protoplanetary disks, and find that the observed data agree with the simulation. Earlier work \citep{Morb09} modeled the gravitational interactions within a population of planetesimals and planetary embryos and finds that to produce a final size distribution consistent with the present day asteroid belt, the initial planetesimal size distribution was dominated by bodies with a minimum size of approximately 100 km, suggesting smaller objects were not present to build planetesimals hierarchically.  

Prior models for planetesimal formation usually assume the hierarchical build-up of kilometre-sized objects from smaller objects via collisions \citep{Kataoka13}. However, this build-up phase would have to occur incredibly efficiently to avoid the aforementioned metre barrier constraints. Thus, the mechanism of planetesimal formation via the gravitational collapse of over-dense clouds of dust pebbles that were generated by the non-linear phase of the streaming instability has become a leading model for this phase of the process of planet formation.

\subsection{Simulating the planetesimal mass distribution}

A primary objective of many studies of planetesimal formation via the SI is to characterize the mass and size distribution of the formed planetesimals \citep{Johansen15,Simon16,Simon17, Schafer17,Abod19, Li19, Gole20}. Such results are useful inputs for models of the evolution protoplanets and planetary cores in the presence of planetesimal disks \citep[e.g.][]{Pollack96}. However, there is still much about simulations of the streaming instability in protoplanetary disks that remains to be understood. 

The SI operates on scales that are a tiny fraction of a protoplanetary disk ($\lesssim 0.01\,AU$), as might be expected of a process that can make $\sim 100$ km-sized bodies. Thus, published 3D numerical simulations have focused on tiny patches in protoplanetary disks. As might be expected, prior work has also focused on regions of parameter space with favourable growth rates which greatly limits the computational expense. In addition, the ubiquitous turbulence and large stopping-distance of dust grains makes the phase space of the dust very complex and difficult to model. This precludes simple adaptive strategies and explains the use of fixed meshes with the associated limits on dynamical range. Thus it is an expensive and ongoing process to explore the full parameter space of dust grain sizes, dust mass, total disk mass, global gas pressure gradient and the role of disk structures. Global disk simulations which resolve the key scales for SI are still far out of reach.

Key questions remain regarding numerical convergence. For example, establishing a minimum planetesimal mass, the detailed properties of the dust density distribution and the turbulent velocity field.  We would also like to investigate the non-linear interactions between the non-linear SI and the full, evolving distribution of grain sizes.  Generally, there is much work to be done in characterizing the non-linear SI, including perturbation growth rates, characteristic length scales, the interaction between newly collapsed planetesimals and dust, the amount of dust converted to planetesimals, the collapse process for individual planetesimals, their resultant properties and the roles of mergers and collisions.

Due to these challenges and the associated computational expense, most studies using 3D simulations to study the planetesimal mass distribution from the SI considered a numerical domain size that was at most 0.2 gas scale heights on a side ($\sim 0.02\,AU$). Thus the impact of larger domains is relatively unstudied. \citet{Yang14} and \citet{Li18} used larger domains in a study of the non-linear SI, but their simulations did not consider gravitational forces between the dust mass, and thus did not follow the development of the non-linear SI all the way to planetesimal formation. \citet{Schafer17} used larger domains that were twice and four times as large in the radial and azimuthal directions and studied the population of planetesimals in the full domain. They constrain parameters of the planetesimal mass distribution in the full domain of the simulation, and they find disagreement in some parameters for the simulations of different sizes, and agreement in other parameters. \citet{Carrera20} used domains with large radial extents to study planetesimal formation via the SI within large-scale, background pressure bumps associated with axisymmetric rings in protoplanetary disks. Larger domains permit new dynamical modes which may impact the planetesimal formation process, but not much research has been done in exploring this impact.

In this paper, we confirm the basic results of \citet{Schafer17}, with a different code and hydrodynamical treatment, using similarly large domains. We expand on their results by running multiple simulations with parameters that are identical but for different random perturbations in initial dust density.  We also briefly examine convergence via enhanced resolution in the largest domain simulation. Through a novel analytical approach we probe the spatial variability in the planetesimal mass distribution and conversion rate of the dust mass to planetesimals throughout the larger domains. We also consider the mass distributions on the scale of the full domain to compare to prior work.

The paper is organized as follows. In Section~\ref{sec:methinit} we outline our methods and parameters of our simulations. In Section~\ref{sec:Results} we describe our methods for analyzing our simulation data and our results. Sections~\ref{sec:cuml} to \ref{sec:diffnum} focuses on the properties of the mass distributions and Section~\ref{sec:pltmass} focuses on the quantifying the total amount of dust that is converted to planetesimals. In Section~\ref{sec:discsumm} we summarize and discuss our results and their impact on the field, as well as future work.

\section{Methods and initial conditions}
\label{sec:methinit}

We model the dynamics of localized portion of a protoplanetary disk, using the shearing sheet approximation \citep{GLB65I} to simulate a local portion of a near-Keplerian, protoplanetary disk with a co-rotating Cartesian frame $(x,y,z)$.  Relative to the central star, the box centre is at $(r, \theta_0, z_0)$ in cylindrical coordinates.  The box is centred on the midplane so that $z_0=0$.  Points within the box are at global coordinates $(r+x,\theta_0+y,z)$.  This approximation neglects the effects of azimuthal curvature in the orbit.

The equations that describe the gas and dust evolution in this non-inertial reference frame are

\begin{align}
 {\frac{\partial \rho_g}{\partial t }} & + \nabla \cdot (\rho_g \bs{u}) = 0  \label{eq:gasmass} \\
 {\frac{\partial \rho_g \bs{u} }{\partial t}} & + 
\nabla \cdot (\rho_g \bs{u}\bs{u}) =  -\nabla P_g \nonumber \\ & + \rho_g \Bigg[-2\bs{\Omega} \times \bs{u} + 2q\,\Omega^2 x\, \hat{\bs{x}}
 -\Omega^2 z\, \hat{\bs{z}} + \mu \frac{\overline{\bs{v}} - \bs{u}}{t_{\text{stop}}} \Bigg] \label{eq:gasmom} \\
 \frac{d \bs{v}_i }{dt} & = 2\bs{v}_i \times \bs{\Omega} + 2q\,\Omega^2 x\, \hat{\bs{x}} - \Omega^2 z\, \hat{\bs{z}} - \frac{\bs{v}_i - \bs{u}}{t_{\text{stop}}} + \bs{F}_g
\end{align}

where $\rho_g$ denotes the gas mass volume density, $P_g$ is the gas pressure, and $\mu \equiv \rho_d/\rho_g$ is the ratio of the local dust mass density to the gas mass density. The velocity of the gas is represented by $\bs{u}$, and the velocity of an individual dust particle is $\bs{v}_i$, where the subscript $i$ identifies the $i$th dust particle.  We use an isothermal equation of state, $P_g = \rho_gc_s^2$, where $c_s$ is the sound speed.

The gas and the dust are coupled together by the terms $\mu (\overline{\bs{v}} - \bs{u})/t_{\text{stop}}$ and $-(\bs{v}_i - \bs{u})/t_{\text{stop}}$ in the gas and dust momentum equations, respectively. The notation $\overline{\bs{v}}$ represents the mass-weighted average velocity of the dust particles in the gas cell (though in our simulations all dust particles have the same mass). The stopping time of the dust particle, $t_{\text{stop}}$, is a timescale that characterizes the rate at which momentum is exchanged between the gas and dust. In the Epstein drag regime \citep{Epstein24}, where the particle size is smaller than the mean free path of the gas, this parameter is given by
\begin{equation}
\label{eq:tstop}
t_{\text{stop}} = \frac{\rho_s s}{\rho_g c_s}
\end{equation}
where $\rho_s$ is the bulk solid density of the particles (approximately $2.6$\,g\,cm$^{-3}$ for silicates \citep{Moore1973})  and $s$ is the radius of the dust grains if we assume they can be approximated with a spherical shape. In protoplanetary disks, the Epstein drag regime applies to dust particles everywhere except the very inner part of the disk \citep{Birnstieletal16}, so we use this drag formalism.

In the local frame described by $(x,y,z)$, which rotates with the Keplerian rotation with the disk, there is a background velocity flow due to differential rotation in the radial direction. The angular velocity is a power law in the disk radius, $\Omega \propto r^{-q}$, and we model Keplerian rotation, where $q=3/2$. In our co-ordinates, the rotation vector is oriented along the $z$-axis, $\bs{\Omega} = \Omega\,\hat{\bs{z}}$, which leads to a background velocity flow given by $(q\Omega x)\hat{\bs{y}}$, where $x$ is the local radial co-ordinate.

\subsection{Numerical methods}
\label{sec:nummeth}

We simulate this system with the public C-version of the \textsc{Athena} hydrodynamics grid code \citep{StoneAth08}. We employ the HLLC Riemann solver to compute the numerical fluxes and the cornered transport upwind (CTU) integrator to evolve the equations in time \citep{StoneAth08, StoneGard09}.  Dust is modeled following \textsc{Athena} \citep{Bai101} with the semi-implicit integrator and the triangular-shaped cloud (TSC) scheme to interpolate particle properties to and from the gas grid.  The gravity solver was modified to include dust self-gravity.  Otherwise, what follows are standard \textsc{Athena} options.

The orbital advection scheme separates the background flow velocity from the fluctuations, leading to a more computationally expedient and accurate algorithm \citep{Masset00, Johnson08, StoneGard10}. Thus the momentum equation for the dust particles which is integrated in our simulations has the background shear flow subtracted, and is of the form
\begin{equation}
\label{eq:dustmomprime}
\frac{d \bs{v}_{i}' }{dt} = 2(v_{iy}' - \eta v_K) \Omega \hat{\bs{x}} - (2 - q)v_{ix}'\Omega \hat{\bs{y}} - \Omega^2 z \hat{\bs{z}} - \frac{\bs{v}_i' - \bs{u}'}{t_{\text{stop}}} + \bs{F}_g
\end{equation}
where $\bs{v}' = \bs{v} - (q\Omega x)\hat{\bs{y}}$ and $\bs{u}' = \bs{u} - (q\Omega x)\hat{\bs{y}}$.

To maintain this shear flow at the radial boundary, our simulations employ shearing box boundary conditions, where the azimuthal ($y$-direction) and vertical ($z$-direction) hydrodynamic boundary conditions\footnote{The boundary conditions are slightly different for the gravity solver, see Section~\ref{sec:selfg}} are purely periodic, and the radial ($x$-direction) boundary conditions are shear periodic \citep[see][]{Hawley95, StoneGard10}. The radial periodic zones move along the $y$-direction with velocities of magnitude $q\Omega L_x$. Once the periodic zones have moved beyond the extent of the computational domain in the $y$-direction, the motion resets and the shear periodic boundary conditions become momentarily purely periodic. The time period for this is given by $t_n = n L_y/(q\Omega L_x)$, where for each $n=0,1,2...$ the radial boundary conditions are purely periodic, and for all intermediate times the are boundaries are not perfectly aligned, according to the shear periodic scheme. Here, $L_x$ and $L_y$ are the extent of the box in the $x$-direction and $y$-direction, respectively.

Another essential component of the streaming instability is large-scale, radial pressure gradients in the gas disk, which has a surface density profile that decreases with radius. This pressure gradient is responsible for maintaining a persistent difference between the radial component of the velocity of the dust and the velocity of the gas. Only the gas feels the radially-outward pointing hydrodynamic force due to this pressure gradient, which causes the gas to orbit at slightly sub-Keplerian speeds \citep{Armitage}. The dust does not feel this force, and orbits at the Keplerian speed. The difference in these radial velocities is small, but it is persistent, which means there is a persistent momentum exchange between the dust and gas via the drag force, hence why this gradient is a key component of the streaming instability \citep{YG05}. 

Including this radial pressure gradient directly in the gas phase within the simulations would create a discontinuity between the inner and outer radial boundaries of the domain. Hence, when including this effect in \textsc{Athena}, \citet{Bai101} approximate the effect of the pressure gradient as a constant force within the shearing box. However, instead of applying an outward radial (positive $x$) force to the gas, a constant inward radial (negative $x$) force is added to the particles. This is the $\bs{F_{\text{grad}}} = -2\eta v_k \Omega \hat{\bs{x}}$ term in equation~\ref{eq:dustmomprime}. The factor $\eta v_k$ measures the amount by which the azimuthal component of the dust and gas is modified from the Keplerian velocity. Given a disk model with a radial pressure profile $P_g \propto r^{-n}$ and an isothermal equation of state, 
\begin{equation}
\label{eq:eta}
\eta = n\frac{c_s^2}{v_k^2}.
\end{equation}
With the gas scale height defined as $H_g \equiv c_s/\Omega$, then $\eta \sim \mathcal{O}(H_g/r)^2$, and in many models of PPDs, e.g. minimum mass solar nebula \citep{Hayashi81}, $H_g/r \sim 0.05$, and a typical value for $\eta$ is $\sim 0.003$. In \textsc{Athena} \citep{Bai101}, this factor $\eta$ in the radial pressure gradient force $\bs{F_{\text{grad}}} = -2\eta v_k \Omega \hat{\bs{x}}$ is parameterized via the dimensionless factor $\eta v_k/c_s$, and the simulations in this study use a value of $\eta v_k/c_s=0.05$ (see Section~\ref{sec:ic}).

\subsubsection{Particle self-gravity}
\label{sec:selfg}

Exploring the creation of bound clumps requires the gravitational acceleration due to dust particles, 
\begin{equation}
\bs{F}_g = -\nabla \Phi_d 
\end{equation}
where the potential due to dust, $\Phi_d$, is the solution of Poisson's equation, 
\begin{equation}
\nabla^2 \Phi_d = 4\pi G \rho_d,
\end{equation}
where $G$ is the gravitational constant.  The TSC interpolation scheme is used to compute the dust density, $\rho_d$ (used for drag and gravity). 

Following prior work \citep[e.g.][]{Simon16}, we neglect the self-gravity of the gas whose local density perturbations are relatively small and also the effect of gravity on gas which is small compared to other forces.  These assumptions can be justified by examining the gaseous \citet{Toomre64} parameter, $Q \equiv c_s \Omega / (\pi G \Sigma) \sim 32$ for our simulations and thus the gas disk is very gravitationally stable (see also equation~\ref{eq:Gtilde} and associated discussion). 

We use the Poisson solver implemented in the public (C-version) of \textsc{Athena} by C.-G. Kim \citep{KimOst17}, with shear-periodic horizontal boundary conditions \citep{Gammie01} and vacuum (open) boundary conditions in the vertical direction \citep{KoyOst09}.  We show tests confirming the correct behaviour of dust with self-gravity in our simulations in Appendix~\ref{sec:app}.

\subsection{Initial conditions \& parameters}
\label{sec:ic}

Our choice for the parameters that control the dust mass, dust grain size, radial pressure gradient, and ratio of gravitational and rotational shear strength are either identical or very similar to choices from previous work \citep{Simon16,Schafer17,Johansen12,Li18,Gole20}. These parameters are summarized in the bottom row Table~\ref{tab:sims} and are defined in this section.

The gas is initialized with a Gaussian profile in the vertical direction
\begin{equation}
\rho_g(z) = \rho_{g,0} \exp \Bigg( {-\frac{z^2}{2H_g^2}} \Bigg)
\end{equation}
where $\rho_{g,0}$ is the gas density in the midplane and $H_g$ is the gas scale height. We set the units of our model so that $\rho_{g,0} = H_g = \Omega = c_s = 1$. 
The dust particle positions are initialized with a random number generator based on a uniform distribution in the $x$-$y$ plane, and a Gaussian profile in the $z$ direction with a scale height $H_d = 0.02 H_g$. The number of particle resolution elements in each simulation is equal to the number of grid resolution elements in the domain. As seen in Table~\ref{tab:sims}, we ran multiple simulations with identical domain sizes and resolutions, each of which labelled with a letter \texttt{a}, \texttt{b}, \texttt{c}, or \texttt{d}. The dust particles in these otherwise identical simulations were initialized with different random number seeds, changing the individual particle positions. This leads to different outcomes in the planetesimal formation process during the non-linear evolution of the streaming instability (explored in Section~\ref{sec:Results}).

The size of the dust grains, $s$, controls the strength of the drag coupling between dust and gas. This sets the dimensionless stopping time,
\begin{equation}
\tau_s \equiv t_{\text{stop}}\Omega.
\end{equation}
In all our simulations, we choose $\tau_s = 0.314$. In terms of orbital periods, $T_{\text{orb}} = 2\pi/\Omega$, we have $t_{\text{stop}}/T_{\text{orb}} \approx 0.05$. The mass of the dust particles is controlled by the ratio of dust mass surface $\Sigma_d$ density to the gas mass surface $\Sigma_g$ density
\begin{equation}
Z \equiv \frac{\Sigma_d}{\Sigma_g}
\end{equation}
and we use $Z=0.02$, a slightly super-solar solid mass ratio. The radial pressure gradient parameter $\eta$ (see equation~\ref{eq:eta}), is parametrized via
\begin{equation}
\Pi \equiv \frac{\eta v_K}{c_s}
\end{equation}
and for this parameter we choose $\Pi = 0.05$. Lastly, the strength of gas self-gravity versus tidal shear is captured by
\begin{equation}
\widetilde{G} \equiv \frac{4\pi G \rho_{g,0}}{\Omega^2}.
\label{eq:Gtilde}
\end{equation}
The value of this parameter sets the relative importance of self-gravity versus tidal shear. Varying $\widetilde{G}$ is equivalent to moving through different radial portions of the disk. For our simulations, as in the fiducial simulation from \citet{Simon16}, we set $\widetilde{G} = 0.05$, equivalent to a Toomre $Q$ of 32. For a disk model where these quantities are power laws in the disk radius $r$, i.e. $\Sigma_g \propto r^{-a}$, $H_g \propto r^{b}$, $\Omega \propto r^{-q}$, then $\widetilde{G} \propto r^{-a-b+2q}$. For $a=1$, $q=3/2$, and, as in the minimum mass solar nebula (MMSN) model \citep{Hayashi81}, $b=5/4$, then $\widetilde{G} \propto r^{3/4}$ and varies with radial position within the disk.

\begin{table}
 \caption{Simulation parameters.}
 \label{tab:sims}
 \begin{tabular}{lcccc}
  \hline
  Run name & 
  \multicolumn{2}{c}{Domain Size} & 
  \multicolumn{2}{c}{Grid Resolution} \\
  
  {} & 
  \multicolumn{2}{c}{$(L_x \times L_y \times L_z)/H_g$} & 
  \multicolumn{2}{c}{$N_{\text{cell}} = N_x \times N_y \times N_z$} \\

  \hline  
  
  \texttt{L02a} & 
  \multicolumn{2}{c}{0.2 $\times$ 0.2 $\times$ 0.2} & 
  \multicolumn{2}{c}{120 $\times$ 120 $\times$ 120} \\
  \texttt{L02b} & 
  \multicolumn{2}{c}{0.2 $\times$ 0.2 $\times$ 0.2} & 
  \multicolumn{2}{c}{120 $\times$ 120 $\times$ 120} \\
  \texttt{L02c} & 
  \multicolumn{2}{c}{0.2 $\times$ 0.2 $\times$ 0.2} & 
  \multicolumn{2}{c}{120 $\times$ 120 $\times$ 120} \\
  \texttt{L02d} & 
  \multicolumn{2}{c}{0.2 $\times$ 0.2 $\times$ 0.2} & 
  \multicolumn{2}{c}{120 $\times$ 120 $\times$ 120} \\
  
  \rule{0pt}{3ex}\noindent
  
  \texttt{L04a} & 
  \multicolumn{2}{c}{0.4 $\times$ 0.4 $\times$ 0.2} & 
  \multicolumn{2}{c}{240 $\times$ 240 $\times$ 120} \\  
  \texttt{L04b} & 
  \multicolumn{2}{c}{0.4 $\times$ 0.4 $\times$ 0.2} & 
  \multicolumn{2}{c}{240 $\times$ 240 $\times$ 120} \\  
  
  \rule{0pt}{3ex}\noindent
  
  \texttt{L08} & 
  \multicolumn{2}{c}{0.8 $\times$ 0.8 $\times$ 0.2} & 
  \multicolumn{2}{c}{480 $\times$ 480 $\times$ 120} \\  
  
  \hline
  \hline
  \rule{0pt}{2ex}\noindent
  
  $N_{\text{par}}/N_{\text{cell}}$ &
  $\tau_s$ &
  $Z$ &
  $\widetilde{G}$ &
  $\Pi$\\
  
  \multicolumn{1}{c}{1} & 0.314 & 0.02 & 0.05 & 0.05\\

  \hline 
 \end{tabular}
\end{table}

\subsection{Simulation domain}
\label{sec:simdom}

\begin{figure*}
	\includegraphics[width=\textwidth]{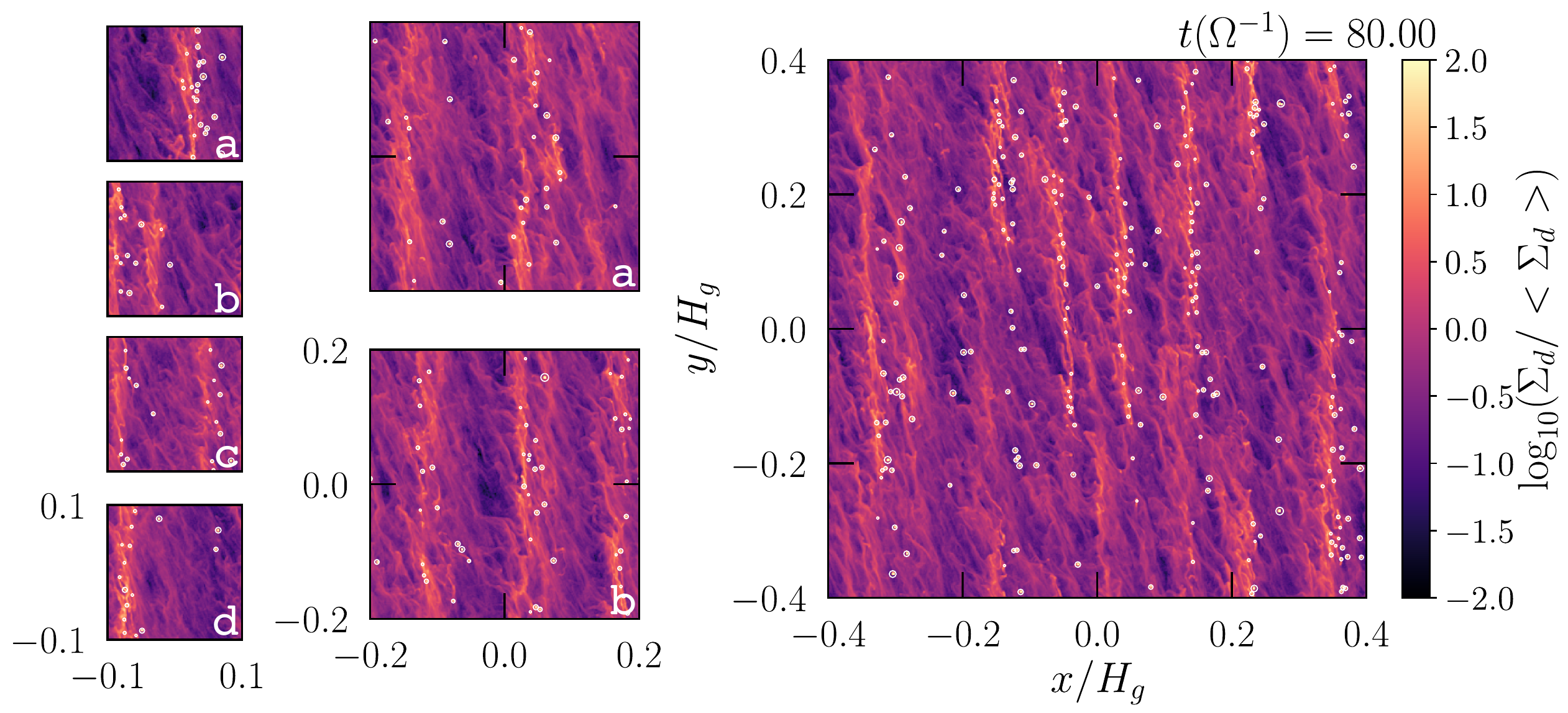}
    \caption{Dust surface density in the $x$-$y$ plane for each of the 7 simulations. The colour represents the logarithm of the dust surface density normalized by the mean dust surface density. Bound planetesimals identified by the group finder are highlighted by the white circles, where the radii of the circles is equal to the Hill radius (equation \ref{eq:RH}). Each snapshot represents the simulation at time $t=80$ in units of the inverse orbital frequency, $\Omega^{-1}$.}
    \label{fig:surfden}
\end{figure*}

In our study, we consider simulation domains of various sizes, as well as multiple runs of simulations with identical physical parameters to investigate the variance planetesimal formation process via the streaming instability. The domain sizes are summarized in Table~\ref{tab:sims}.

We employ simulations with $L_x = L_y = L_z = 0.2$, as well as $L_x = L_y = 0.4,\ L_z = 0.2$ and $L_x = L_y = 0.8,\ L_z = 0.2$, where all above lengths are in units of the gas scale height, $H_g$. We introduce a shorthand for the simulations with the previously described domain sizes, and refer to them as \texttt{L02}, \texttt{L04} and \texttt{L08}, respectively.

We maintain an equivalent numerical resolution (in terms of cells per length) between runs. In our smallest domains, the \texttt{L02} runs, which matches the size of the domains from \citet{Simon16}, we use a moderate resolution of $(N_x,N_y,N_z)=(120,120,120)$. This results in cubic resolution elements in our simulation grids, with a side length of $0.2 H_g/120 \approx 0.00167 H_g$. We maintain this resolution in our larger simulations, hence the \texttt{L04} runs have $(N_x,N_y,N_z)=(240,240,120)$ and the \texttt{L08} runs have $(N_x,N_y,N_z)=(480,480,120)$.

We note that, according to \citet{Simon16}, for these dust parameters our resolution of $\sim 0.001667 H_g$ is sufficient to adequately sample the planetesimal distribution, typically providing several planetesimals per \texttt{L02} sized box. At higher resolutions, the dust particles can collapse to smaller length scales because gravity is discretized at the grid cell scale, and thus smaller mass planetesimals can be formed, and a greater number of planetesimals overall. At lower resolutions, only a few planetesimals per \texttt{L02} box can form.

While the ratio of the dust-to-gas mass surface density is $Z=0.02$, the ratio of the midplane dust mass density and dust gas density, given by,
\begin{equation}
    \frac{\rho_{d,0}}{\rho_{g,0}} \equiv \frac{\Sigma_d}{\Sigma_g}\frac{H_g}{H_d} = Z\Bigg(\frac{H_d}{H_g}\Bigg)^{-1},
\end{equation}
is actually rather high once the dust settles to the midplane. The ratio $H_d/H_g$ approaches $\sim 0.05$, which gives $\rho_{d,0}/\rho_{g,0}\sim 0.4$, approaching unity. Also, as shown in the next section, the relationship between the total dust mass and the total gas mass in the simulation domain is $M_{\text{dust,T}} = 0.25 M_{\text{gas,T}}$. This is because the vertical extent of the box is $0.2 H_g$, which excludes a significant portion of the gas mass in this small patch of the protoplanetary disk, while all the dust mass in the vertical dimension is included within the domain (recall $H_{d,0}=0.02H_g$).

\subsection{Physical unit conversion}
\label{sec:physunit}

Following \citet{Simon16} and \citet{Johansen12} we convert to physical units by considering a mass unit given by $M_0 = \rho_{g,0}{H_g}^3$, and then use the MMSN model \citep{Hayashi81} for the gas scale height as a function of disk radius, $H_g(r) \sim 0.033 (r/\text{AU})^{5/4}$. With $r=3$ AU, we have $M_0 = 6.7\times 10^{26}$ g. For our smallest (\texttt{L02}) boxes, the total amount of gas in the box is $M_{\text{gas,T}} \approx 0.008 M_0$. With $\Sigma_g = \sqrt{2\pi}\rho_{g,0}H_g$, $\Sigma_d = M_{\text{dust,T}}/(L_xL_y)$, we have
$M_{\text{dust,T}} = \sqrt{2\pi}(L_x/H_g)(L_y/H_g) Z M_0$. Again, for the \texttt{L02} boxes, this gives $M_{\text{dust,T}} \approx 0.002 M_0 = 0.25 M_{\text{gas,T}}$ and with the conversion for $M_0$ to physical units, assuming a global disk radius of $r=3$ AU, the total mass of dust in the \texttt{L02} boxes under these assumptions is $M_{\text{dust,T}} = 1.34 \times 10^{24}\ \textrm{g} \approx 1.5 M_{\text{Ceres}}$.

With the same MMSN prescription for $H_g(r)$ as above, $0.2 H_g$ (the side length of our smallest domain) converts to $\sim 0.025$ AU if we place the simulation box at $r=3$ AU. At the same radius, our resolution unit of $\sim 0.00167 H_g$ converts to $\sim 2\times 10^{-4}$ AU, or $32,000$ km.

\subsection{Computational resources}
Every simulation in this study was integrated to at least $t=200\Omega^{-1}$ in \textsc{Athena}. The number of CPU hours used to integrate to $t=200\Omega^{-1}$ was $\sim$3500 for each \texttt{L02} simulation, $\sim$8200 for each \texttt{L04} simulation, and 27400 for the \texttt{L08} simulation. All simulations were run on the ComputeCanada Niagara cluster.

\subsection{Planetesimal mass distribution characterization}
\label{sec:massdist}

In this section we describe the methods we used to quantify the mass distribution of planetesimals formed in our simulations.   The cumulative mass distribution, $N_>(m_p)$, is the number $N$ is the number of planetesimals of greater or equal mass than $m_p$.  Following \citet{Simon17}, we estimate the differential mass distribution via, 
\begin{equation}
\label{eq:dNdMcalc}
\frac{dN}{dm_p}\Bigg|_i = \frac{2}{m_{p,i+1}-m_{p,i-1}}.
\end{equation}
where $i$ denotes the $i$th planetesimal ranked in increasing mass.   We use the maximum likelihood estimator (MLE) of \citet{Clauset09} to estimate the power-law index $p$ such that $\frac{dN}{d\,m_p} \propto m_p^{-p}$.  This gives,
\begin{equation}
\label{eq:pcalc}
p = 1 + n \Bigg[ \sum_{i=1}^n \ln\Bigg( \frac{m_{p,i}}{m_{p,\text{min}}} \Bigg) \Bigg]^{-1},
\end{equation}
where $n$ is the number of planetesimals in the set of planetesimal masses, $\{m_{p,i}\}$, and $m_{p,\text{min}}$ is the minimum planetesimal mass in the set. The error in the estimate for $p$ is,
\begin{equation}
\label{eq:sigcalc}
\sigma = \frac{p-1}{\sqrt{n}}.
\end{equation}

Other studies \citep{Schafer17, Li19} characterized the mass distribution with a variety of functions that contain more parameters, including some that combined a power-law fit with an exponential cut-off. Since we use only moderate resolution and thus have lower planetesimal numbers than the high-res simulations from \citet{Simon16}, we choose to only fit our data with a single power law.

\subsection{Group finding}
\label{sec:grpfnd}

We employ the group finding algorithm \textsc{SKID} \citep{StadelSKID} to identify gravitationally bound clumps in our particle data, which we refer to as planetesimals in our study.

The Hill radius, $R_H$, characterizes the roughly spherical region where a planetesimal's gravity dominates over shear \citep{Armitage}.  This radius can be expressed as,
\begin{equation}
R_H = \Bigg( \frac{m_p G}{3\Omega^2} \Bigg)^{1/3},
\label{eq:RH}
\end{equation}
which gives the Hill density for a planetesimal with mass $m_p$, 
\begin{equation}
\rho_H \equiv \frac{3}{4\pi}\frac{m_p}{R_H^3} = 9 \frac{\Omega^2}{4\pi G}.
\end{equation}
The SKID algorithm computes a mass density estimate on the dust particle data, and we consider any clumps with densities above $\rho_H$ and with a sufficiently large mass $m_p$ so that the Hill radius for that clump is greater than the width of the hydrodynamic grid cell, $\Delta x = L_x/N_x$. These are the same conditions used in \citet{Li19} and \citet{Gole20}, who likewise employed a clump finding algorithm on the dust particle data to identify planetesimals. We note that the results of our study are not sensitive to these cut-offs as most of the identified planetesimals are massive enough that their Hill radius $R_H$ is much larger than $\Delta x$, and the densities of the particles in these clumps are well clear of $\rho_H$, confirming that these particles are unambiguously gravitationally bound.

\section{Planetesimal mass distribution}
\label{sec:Results}

In this section we examine the variability in the formation of planetesimal via the streaming instability.  We explore this via simulations with domains of varying sizes and re-runs of otherwise identical simulations with different random seeds used to distribute the dust particles (see Section~\ref{sec:ic}).

Figure~\ref{fig:surfden} shows the dust surface density in the $x$-$y$ plane for each of our simulations at $t=80 \Omega^{-1}$. We choose to present the dust surface density and perform our mass distribution analyses at $t=80 \Omega^{-1}$ because at this time, enough planetesimals have formed to sample the distribution well, but this is also before planetesimals have grown substantially\footnote{In Section~\ref{sec:discsumm} we discuss how the cross-sections of the bound dust objects in the simulations in this study (and all similar studies) are unrealistically large, and how this impacts the mass distribution over time.}. The planetesimals in Figure~\ref{fig:surfden} are highlighted with white circles.  Visually, it is clear that the distribution of dust varies significantly amongst the simulations with the same domain size and different random seeds.  For the larger domain runs (such as \texttt{L08}), regions that have the same area as an entire \texttt{L02} run may contain many more or many fewer planetesimals at the same state of evolution.\subsection{Cumulative number distributions}
\label{sec:cuml}

\begin{figure}
	\includegraphics[width=\columnwidth]{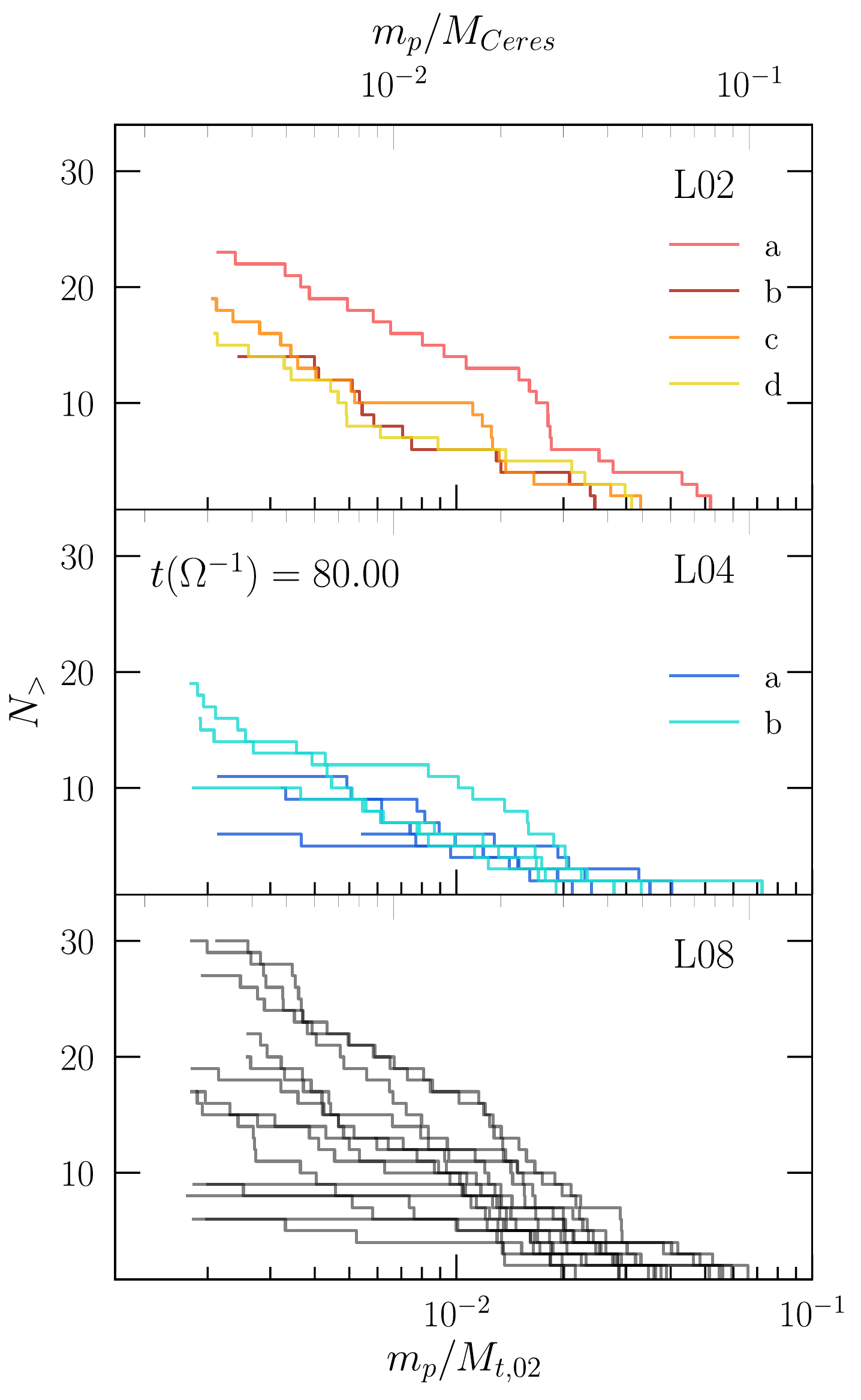}
    \caption{Cumulative number distributions of the planetesimal mass at time $t=80 \Omega^{-1}$. For the \texttt{L04} and \texttt{L08} data, the simulation domains have been subdivided into smaller boxes equivalent in size to the \texttt{L02} domains (see Section~\ref{sec:cuml}). The data represents the planetesimal distribution at the simulation time $t=80 \Omega^{-1}$, the same snapshot considered in Figures~\ref{fig:surfden} and \ref{fig:diffdists}. The planetesimal masses are given in units of the total mass of the dust in an \texttt{L02}-sized domain on the bottom $x$-axis and the mass of Ceres on the top $x$-axis (see Section~\ref{sec:physunit} for physical unit conversions).}
    \label{fig:cumldists}
\end{figure}

For these data, we subdivide the larger simulations (\texttt{L04a}, \texttt{L04b}, \texttt{L08}) into regions with the same area as the \texttt{L02} runs.  The cumulative number distributions for each sub-region are shown as separate lines in Figure~\ref{fig:cumldists}. Explicitly, there are 4 such sub-domains for each \texttt{L04} run and 16 for \texttt{L08}.  

Figure~\ref{fig:cumldists} demonstrates the large variability in the cumulative number distribution for the planetesimal masses at $t=80 \Omega^{-1}$ in these equal area regions. At the mass $m_p/M_{t,\textrm{02}} = 0.03$, the spread in the number of planetesimals within the different \texttt{L02} simulations is 14 to 22, and in the \texttt{L04} simulations the spread is 6 to 14, and in the \texttt{L08} the spread is 6 to 29. This spread--most easily seen in the \texttt{L08} simulation, which represents largest total area with 16 \texttt{L02}-sized boxes--demonstrates the variable behavior in the planetesimal formation process via the streaming instability that is not represented well by even a few \texttt{L02} simulations.

There is also variation in how these planetesimals are distributed in mass. There are many planetesimals between $0.02\ M_{\text{Ceres}}$ and $0.03\ M_{\text{Ceres}}$ in the \texttt{L02a} run and between $0.01\ M_{\text{Ceres}}$ and $0.02\ M_{\text{Ceres}}$ in the \texttt{L02c} run, but the other \texttt{L02} runs do not have many planetesimals at these masses. This trend is observed in the samples of \texttt{L02}-sized domains within the larger boxes as well.

\subsection{Differential number distributions}
\label{sec:diffnum}

\begin{figure}
	\includegraphics[width=\columnwidth]{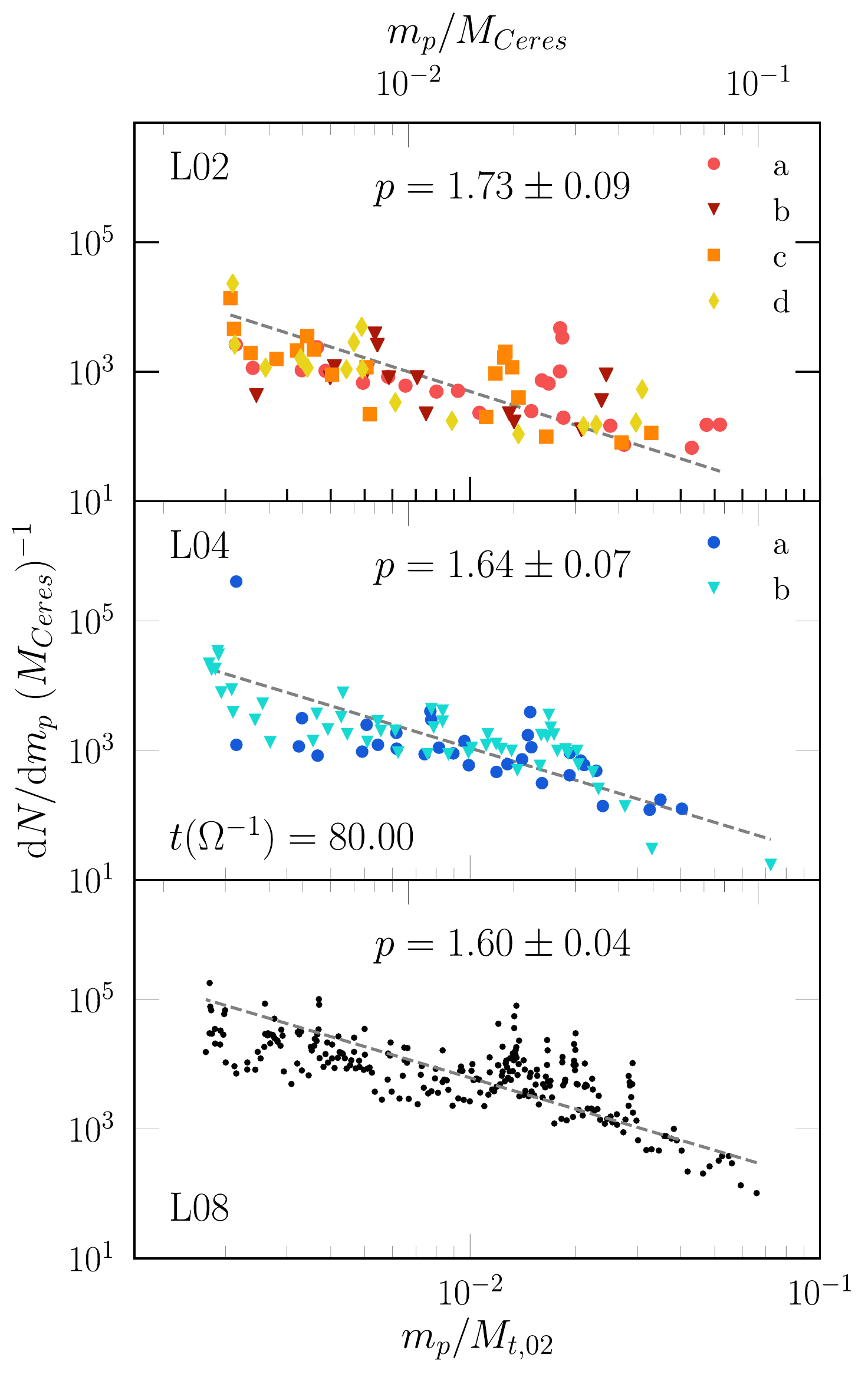}
    \caption{Differential number distributions of the planetesimal mass, computed according to equation~\ref{eq:dNdMcalc}. The data for each of the 7 simulations is plotted individually. The grey dashed lines are power-law fits, where the slopes are calculated according to equation~\ref{eq:pcalc}. The calculation for the fit in the \texttt{L02} panel includes every data point for all four simulations, and the fit in the \texttt{L04} panel includes every data point in the two simulations. The data represents the planetesimal distribution at the simulation time $t=80 \Omega^{-1}$, the same snapshot considered in Figures~\ref{fig:surfden} and \ref{fig:cumldists}. The planetesimal masses are given in units of the total mass of the dust in an \texttt{L02}-sized domain on the bottom $x$-axis and the mass of Ceres on the top $x$-axis. See Section~\ref{sec:physunit} for details on the conversion to physical units (see Section~\ref{sec:physunit} for physical unit conversions).}
    \label{fig:diffdists}
\end{figure}

\begin{figure}
	\includegraphics[width=\columnwidth]{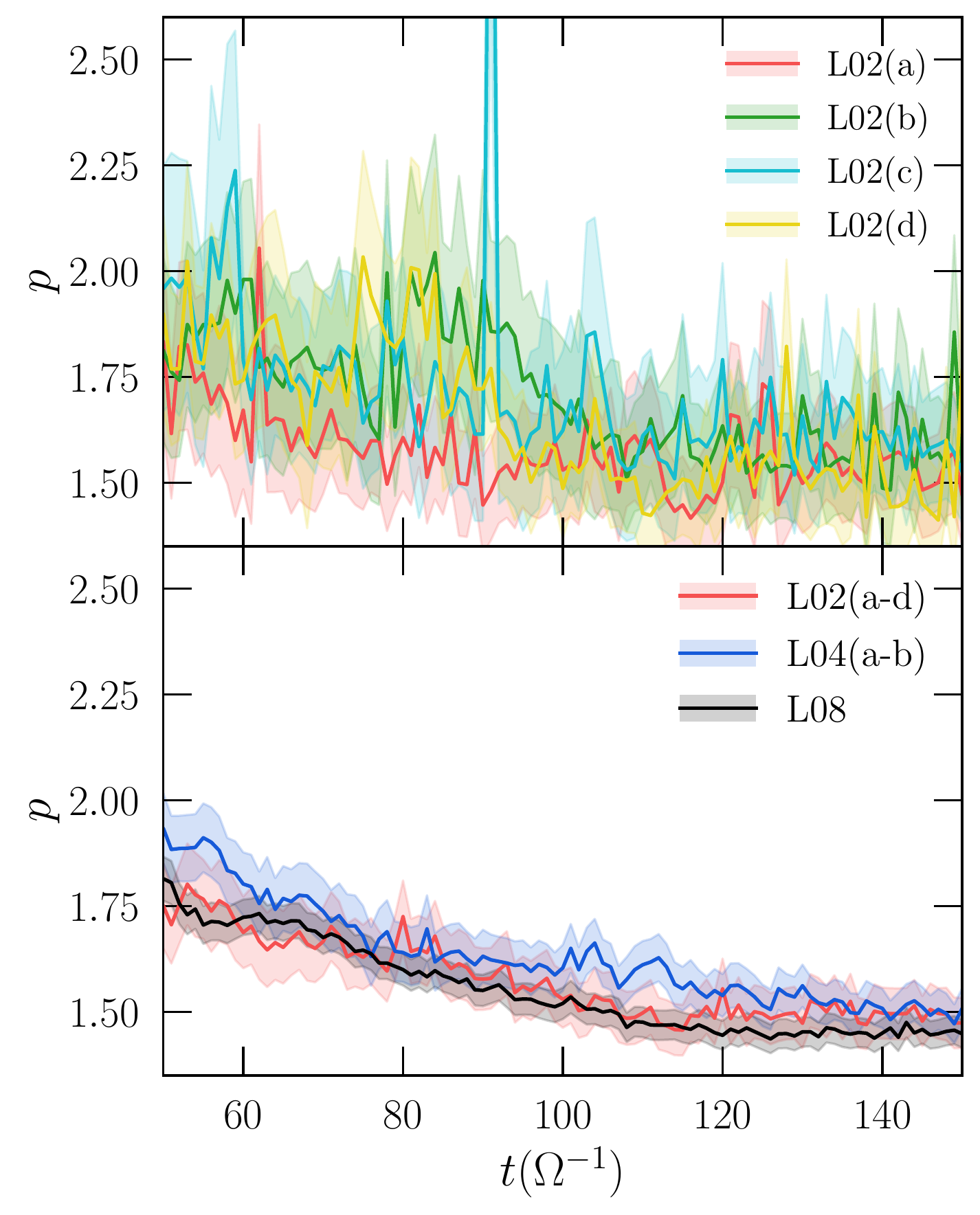}
    \caption{The slope of the power-law fit to the $\text{d}N/\text{d}m_p$ mass distributions over time. The power-law index $p$ is calculated according to equation~\ref{eq:pcalc}. \textit{Top.} The data for the \texttt{L02} simulations. \textit{Bottom.} The data for all 7 simulations. As in Figure~\ref{fig:diffdists}, when computing the power-law index, every data point in all of the \texttt{L02} simulations is included in the calculation for $p$, and the same goes for the \texttt{L04} data. \textit{Both.} The shaded region represents $\sigma$, the error in $p$, calculated according to equation~\ref{eq:sigcalc}.}
    \label{fig:p}
\end{figure}

Figure~\ref{fig:diffdists} shows the differential mass distributions, estimated as described in~\ref{sec:massdist}. Each symbol in the top panel represents $dN/dM$ for just one of the four \texttt{L02} simulations.  However, the indicated power-law index $p$ was computed with all four runs.  The same procedure was used for the \texttt{L04} runs in the middle panel.

We find power-law indices of $p_{02} = 1.73 \pm 0.09$, $p_{04} = 1.64 \pm 0.07$, $p_{08} = 1.60 \pm 0.04$ for the different domain sizes. The decreasing uncertainty reflect the larger total area. Within this modest uncertainty, the different cases agree with each other and are also generally in agreement with values reported in \citet{Simon16}, \citet{Simon17} and \citet{Johansen15}.

The mass distribution of the planetesimals changes over the course of the simulations and this is reflected in the indices as shown in Figure~\ref{fig:p}. When considering the small domain simulations individually, as in the top panel, there is a lot of variance in the value of $p$, typically ranging from $1.5$ to $2.0$, and upper and lower limits exceeding that.  This partly reflects the total numbers in each sample being in the range of 10-30 at the chosen resolution. There is a general trend to less variation at later times and smaller $p$ values.

In the bottom panel, when the larger domains are considered and the planetesimal populations from the multiple \texttt{L02} and \texttt{L04} runs are combined, there is much less variance in the value for $p$. The steady, decreasing trend with time is readily apparent. At $t=50\Omega^{-1}$, when enough planetesimals have formed to compute a reliable value for $p$, the values range between $1.6$ and $2.0$ across the different sized simulations, and well after planetesimals have formed, at $t=150\Omega^{-1}$, the values are between $1.4$ and $1.6$. A decrease in $p$ represents a shift towards fewer and more massive planetesimals at late times.

A trend toward larger masses with time is somewhat expected.  However, a clear demonstration of this trend has not been demonstrated in previous studies.  In Figure 3 from \citet{Simon17}, the authors show data for $p$ over time in their simulations, but only over a relatively narrow range of time\footnote{Our physical dust parameters very closely match the simulation from the middle panel of their Figure 3.}.  Similarly, \citet{Schafer17} show how the values of their fit parameters change over time, but also only for a narrow window. This decrease in distribution fit parameters emphasizes that care is needed when attempting to extract a single value for the power-law index or a single set of parameters that describes the mass distribution of planetesimals formed by the streaming instability. The mass distribution is transient and should be expected to evolve indefinitely, albeit as a slowing rate, particularly when a larger simulation domain provides for more material as shown in the next section.

\subsection{Total mass of dust in planetesimals and the onset of planetesimal formation}
\label{sec:pltmass}

\begin{figure}
	\includegraphics[width=\columnwidth]{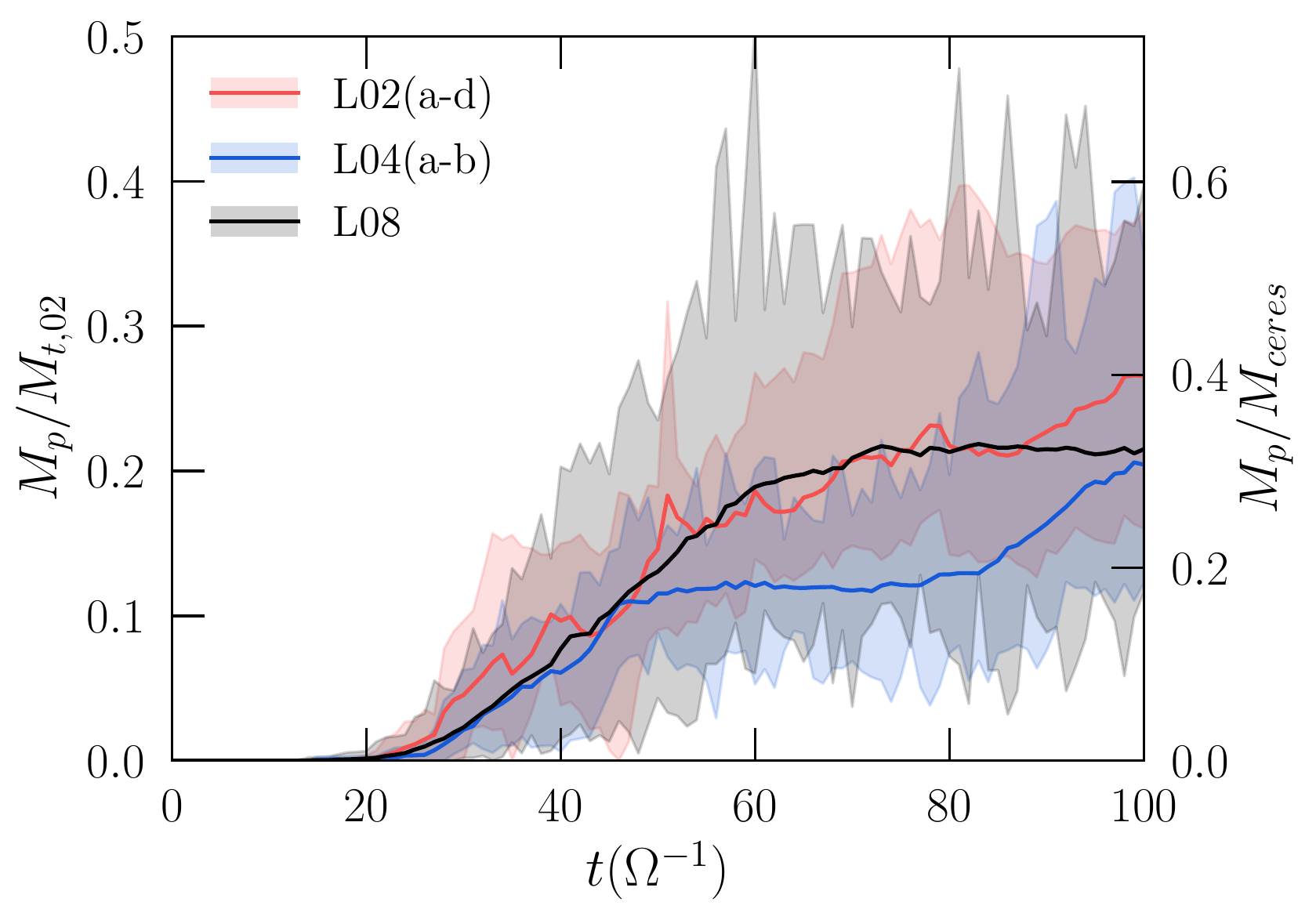}
    \caption{The total mass of the planetesimals $M_p$ in the simulations over time. For the \texttt{L04} and \texttt{L08} data, the simulation domains have been divided into smaller boxes equivalent in size to the \texttt{L02} domains (see Section~\ref{sec:cuml}). For the \texttt{L02} data, all four of the small domain simulations are considered simultaneously, and the \texttt{L04} data considers both of the intermediate sized simulations. The solid line represents the average mass of planetesimals for all the \texttt{L02} or \texttt{L02}-sized boxes, and the shaded region is bounded by the maximum and minimum values in this set. The planetesimal masses are given in units of the total mass of the dust in an \texttt{L02}-sized domain on the left $y$-axis and the mass of Ceres on the right $y$-axis. See Section~\ref{sec:physunit} for details on the conversion to physical units.}
    \label{fig:Mplt}
\end{figure}

\begin{figure}
	\includegraphics[width=\columnwidth]{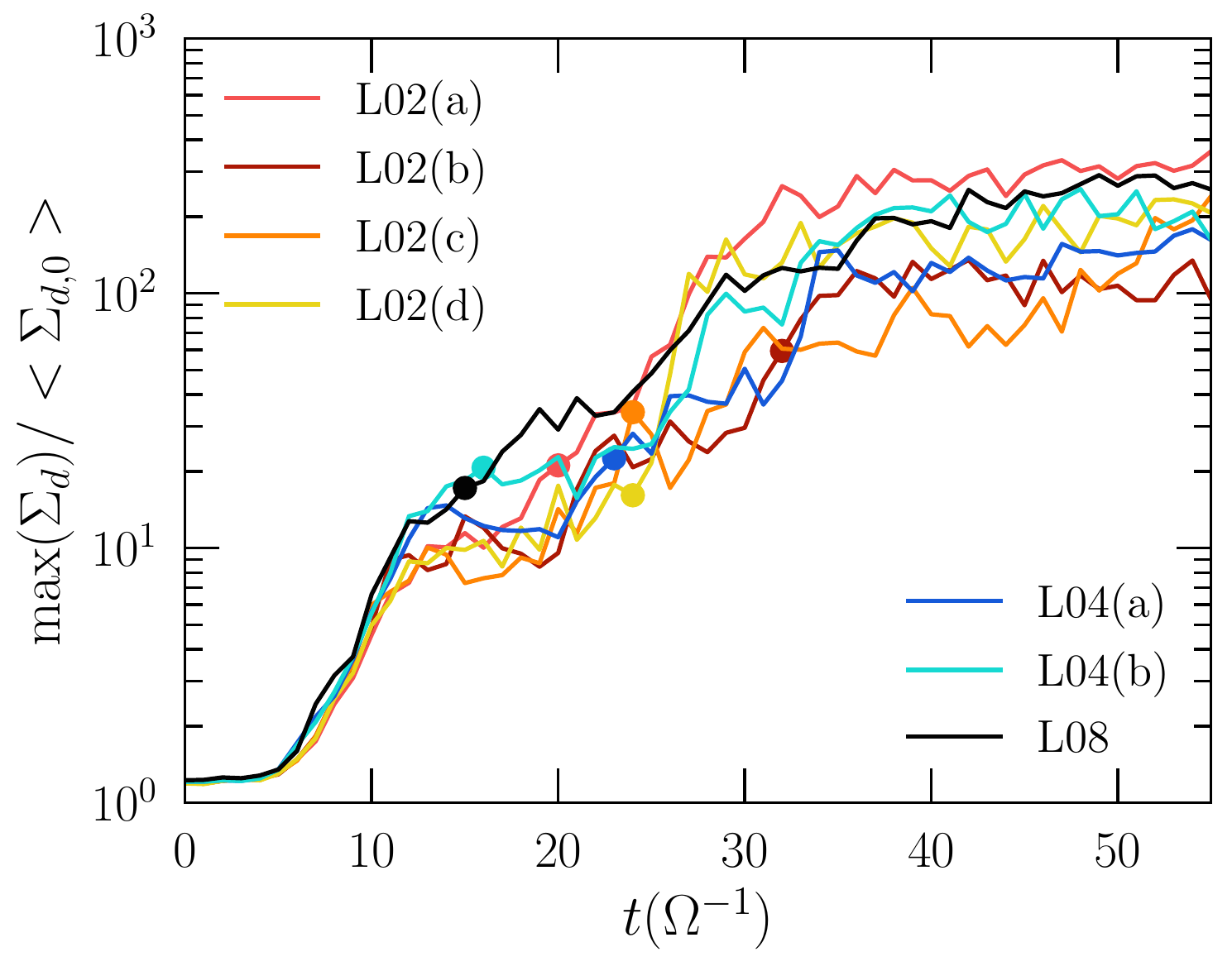}
    \caption{Maximum value for the dust surface density in the $x$-$y$ plane of the 7 simulations over time, normalized by the mean dust surface density. The circles represent the point in time where each simulation first formed planetesimals.}
    \label{fig:Dsig}
\end{figure}

Figure~\ref{fig:Mplt} shows the total mass of the planetesimals in the simulations over time. As in Figure~\ref{fig:cumldists}, the larger domain simulations are divided into smaller sub-domains with the same area as the \texttt{L02} runs. The \texttt{L08} run shows the largest variance in these data. After $t=40\Omega^{-1}$, the spread in the total dust mass in planetesimals in any of the sub-domains from the \texttt{L08} run spans $5\%$ to $45\%$ of the total mass of dust in a single sub-domain. The data from the \texttt{L02} and \texttt{L04} runs generally fit within the maximum-minimum bounds of the \texttt{L08} run. Once again we note that simply re-running these simulations with a different random seed  leads to significantly different consequences for planetesimals formation, shown here directly by the wide spread denoted by the red shaded region that is quite similar to the region-to-region variation in the larger domains.

Figure~\ref{fig:Dsig} shows the maximum value of the dust surface density, $\text{max}(\Sigma_d)$, in the $x$-$y$ plane over the course of all 7 simulations.  Before approximately $t \sim 10\Omega^{-1}$, all simulations evolve quite similarly, however, between about $t \sim 10\Omega^{-1}$ to $t \sim 20\Omega^{-1}$, the larger domain simulations have the highest values of $\textrm{max}(\Sigma_d)$. After this time highly turbulent motions are present in the dust dynamics and the chaotic evolution of the dust density leads to diverging tracks.    

The point where planetesimals first form in each simulation (denoted by the circles in Figure~\ref{fig:Dsig}) spans a range of $t=15\Omega^{-1}$ to $35\Omega^{-1}$. This is another representation of the non-linear nature of the streaming instability: even among nearly identical simulations, the dust surface density can evolve differently, which affects the timing for planetesimal formation. Also, the first formation of planetesimals tends to occur earlier in the bigger domains. This is likely related to the observation that the value of $\textrm{max}(\Sigma_d)$ is higher in the larger domains from $t \sim 10\Omega^{-1}-20\Omega^{-1}$. Planetesimal formation requires large over-densities, and the simulations that first reach dust densities sufficient for gravitational collapse will be the first to form planetesimals. The large domain simulations can more quickly reach high dust over-densities because large scale dynamical modes can enable a faster growth to more extreme local density maxima. The influence of these large scale modes can also be seen in the variation in the spatial distribution of the planetesimals at $t=80\Omega^{-1}$ in Figures~\ref{fig:surfden} and \ref{fig:cumldists}. The smaller \texttt{L02} domains cannot represent the large scale modes available in the \texttt{L08} domains. We will quantify and discuss the presence of these large scale modes in an upcoming paper in this series.

\section{Summary \& Discussion}
\label{sec:discsumm}

In this study we used 3D simulations of patches of protoplanetary disks to study the formation of planetesimals from the gravitational collapse of dust over-densities generated by the streaming instability. We employ simulations that use larger domains than most studies and higher resolution than a study that used similar sized domains. Also, we re-run simulations with identical physical parameters except for the randomized placement of the dust particles--a novel approach for these kinds of simulations. Both the larger domains and re-run simulations allow us to probe the variability in the population of planetesimals which is caused by the non-linear nature of the streaming instability. Our main results are as follows:

\begin{enumerate}
    \setlength{\itemsep}{5pt}
    \item The cumulative number distribution for the planetesimal mass in any of the single \texttt{L02} domains (which represent the maximum domain size used by most similar studies) or \texttt{L02}-sized sub-domains within the larger simulations exhibits large variability. The re-run \texttt{L02} simulations exhibit a spread in the total number of planetesimals that ranges from 14 to 22, and this spread is 6 to 29 in the sub-domains within the \texttt{L08} simulation. That is, there is greater variability in the planetesimal distribution in the larger domain simulations than the smaller domains. The number of planetesimals at specific masses is also highly varied within the different \texttt{L02} or \texttt{L02}-sized domains.
    \item Variability in the planetesimal formation process can also be seen in the total mass of dust converted to planetesimals within these domains. In the case of re-run \texttt{L02} simulations, the mass conversion rate to planetesimals varies between 5 and 25\%, and within the domain of the \texttt{L08} simulation this conversion is between 5 and 45\%. Spatial variability in the planetesimal formation process has not previously been reported in other studies.
    \item In our study we characterize the differential number mass distribution of planetesimals with a single parameter: a power-law index. The value of this parameter is consistent across our three different choices of domain size when all planetesimals for each domain size are considered together, and our values as consistent with the index measured by other studies. However, we find these indices decrease over time, by as much as $\sim 10\%$ over the course of several orbits. This is representative of the planetesimal population becoming more top-heavy, i.e. the largest planetesimals disproportionately increase in mass over the course of the simulation. Thus, identifying a single choice of parameters that describes the mass distribution may be intrinsically difficult in our simulations and similar set-ups.
    \item The dust surface density in the radial-azimuthal plane in the \texttt{L08} simulation displays box-scale structure in the azimuthally oriented filaments. In this large domain, the filaments do not span the full azimuthal extent as in the smaller domain simulations. The distribution of planetesimals is also clearly unevenly distributed in the azimuthal directions. This implies large-scale dynamical modes which are not present in the small domains are contributing to the highly variable planetesimal formation process observed in the \texttt{L08} simulation. In subsequent work, we intend to quantify these larger-scale modes and their role.
    \item The maximum surface density grows quicker and planetesimals form earlier in larger domains simulations. This suggests an active role for larger-scale dynamical modes that exists in the larger domains but cannot be represented by the smaller domains.  Again, we defer a detailed exploration of large-scale modes to a upcoming work where we will consider filament evolution leading up to planetesimal formation.
\end{enumerate}

\subsection{Ongoing challenges and future work}
When characterizing the planetesimal mass distribution in our simulation, and in all studies that employ similar techniques, a fundamental issue arises due to limited computational power. At the resolution in our study, the minimum length scale that is resolved, i.e. the cell-size, converts to approximately $\sim 30,000$ km in physical units (see Section~\ref{sec:physunit}). The gravitational force is discretized at this length scale, meaning this is the smallest sized bound object that can represented in our simulation.  We should aim to probe kilometre and tens of kilometres length scales: the true length scale of planetesimals, extrapolated from observations of asteroids and Kuiper Belt objects.  If we kept the same domain sizes from this study, we would require some 1000 times better resolution, or 1000 times more grid points in each dimension. This is beyond the reach of current computational capabilities. Our conclusion then is the smallest planetesimals in our study (and all studies of this variety) do not accurately represent what we would expect to be the true smallest planetesimal mass in nature. \citet{Simon16} use higher resolution simulations in their study and the minimum planetesimal mass in that study is not converged. This means that the low-mass end of the planetesimal mass distribution in such studies is still an open question. The minimum size of the planetesimals is an important parameter in studies that model the interior evolution of the planetesimals to constrain the planetesimal formation timescales in the early Solar system \citep{Lich18}.

A second effect of the large grid cell size is the enhancement of planetesimal-planetesimal interactions such as mergers and planetesimal-disk interactions such as the accumulation of dust material post-formation, compared to what would occur in nature. As mentioned in our summary point (iii) above, we observe that the mass distributions become increasingly top-heavy over time, but this phenomenon is likely more pronounced in this and all similar work due to artificially large interaction cross-sections. Planetesimals could accrete mass after formation, but not with effective collisional cross-sections of $\sim\,1$ billion km$^2$. To combat this issue, \citet{Gole20} use a clump-tracking algorithm to identify planetesimal masses at the moment they are formed in their simulation. This probes the ``birth'' mass distribution, and avoids including planetesimals that may have grown artificially large. \citet{Johansen15} and \citet{Schafer17} replace bound dust objects with sink particles but find this does not substantially change the mass distribution. The objective of our study, which used moderate resolution, was not to definitely explore the planetesimal mass distribution itself, so we do not employ these more advanced techniques. Instead, we study how these outcomes vary due to larger domain simulations and across a sample of re-run simulations. Our methods are sufficiently accurate for those purposes and illustrate the impact of domain size and intrinsic variation.

Characterizing the azimuthally-oriented dust filaments formed by the non-linear SI (readily visible in Figure~\ref{fig:surfden}) will be essential for establishing a broader understanding of planetesimal formation via the SI.  
These filaments are where dust over-densities become large enough to gravitationally collapse, hence they comprise the material reservoirs for planetesimal formation. Key characteristics include their radial width, and radial separation. The non-linear physics that produces these filaments makes a priori predictions from analytical theory difficult.  A few studies have empirically investigated these length scales \citep{Yang14, Gerbig20}.  Of particular interest is whether scales significantly larger than typical simulation boxes could affect filaments and consequent planetesimal formation.  In a subsequent paper in this series, we will explore the origin and impact of characteristic dust filament lengths scales and the role of large-scale dynamical modes.

\section*{Acknowledgements}

These simulations were performed on the Niagara system operated by SciNet and Compute Canada. JW thanks NSERC for funding support.

\section*{Data availability}
The data underlying this article will be shared on reasonable request to the corresponding author.




\bibliographystyle{mnras}
\bibliography{library} 




\appendix

\section{Self-gravitating shearing wave test}
\label{sec:app}

\begin{figure}
	\includegraphics[width=\columnwidth]{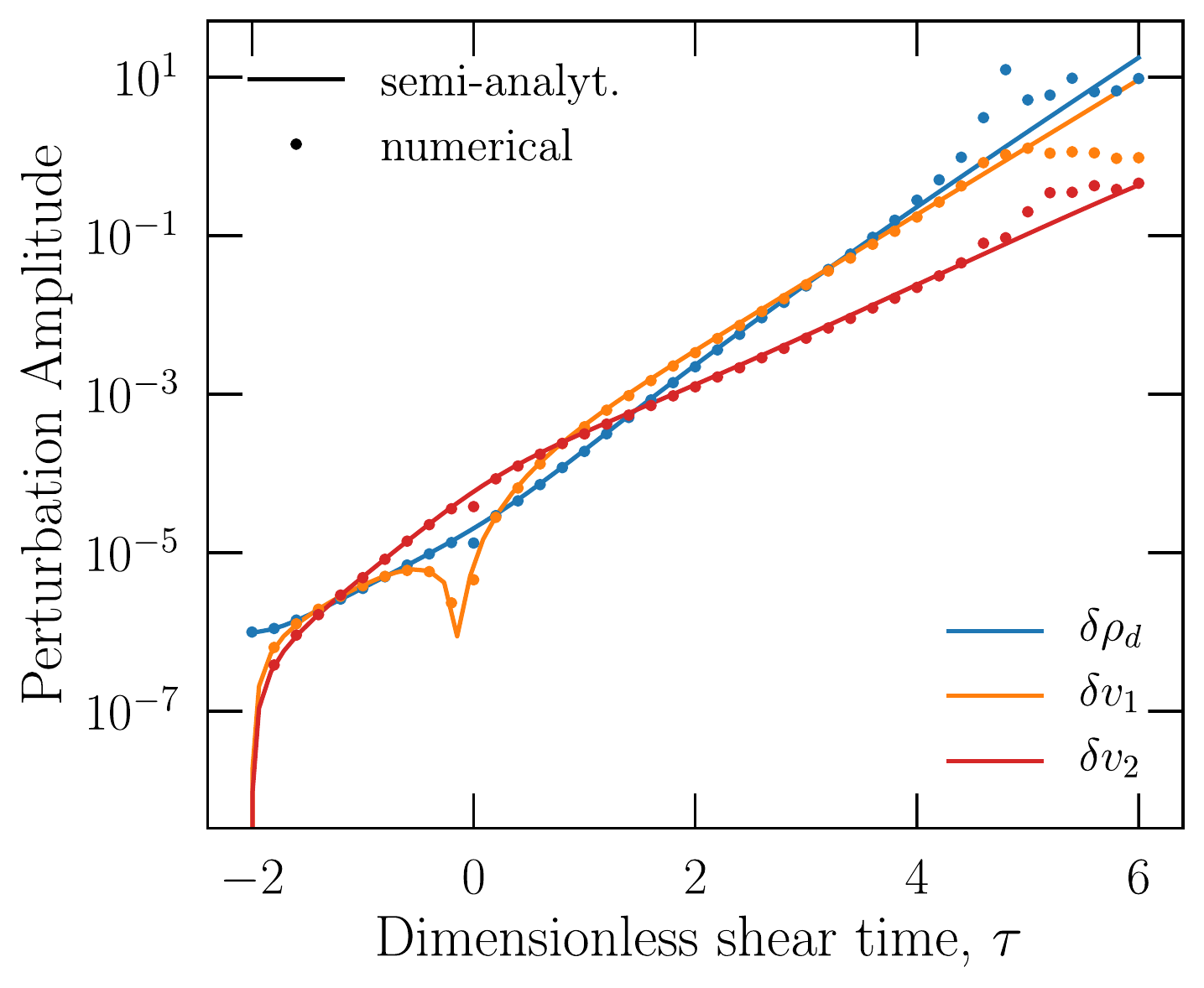}
    \caption{Evolution of the amplitudes of the self-gravitating dust fluid from the shearing wave test. The solid lines represent the evolution from a semi-analytical Runge Kutta integration, and the dots represent the evolution from the numerical (\textsc{Athena}) integration. The time axis is units of the dimensionless shearing time parameter $\tau$.}
    \label{fig:appshearwave}
\end{figure}

To test our implementation of self-gravity applied to the dust particles, we used the shearing wave test from Section 2.2.2 of \citet{Simon16} and Section 1.3.1 of the Supplementary Information from \citet{Johansenetal07}, which is based on methods from \citet{GLB65I}. In this set-up, the initial condition is a plane wave perturbation in the $x$-$y$ (radial-azimuthal) plane and uniform properties in the $z$ direction, and the amplitude of the wave is small compared to the background follows so that the evolution of the amplitude can be described by a linear approximations to the hydrodynamic equations. As in \citet{Simon16} and \citet{Johansenetal07}, we compare the evolution of the amplitudes from the numerical integration in \textsc{Athena} to a semi-analytical Runge-Kutta integration of the amplitudes computed using the \texttt{solve\textunderscore ivp} routine from the \texttt{scipy.integrate} module of SciPy ver. 1.1.0 \citep{Scipy2020}. 

The numerical integration used the shearing box configuration in \textsc{Athena} with purely periodic boundary conditions in $y$ and $z$ and shear-periodic boundaries conditions in $x$. Also, to isolate the influence of the self-gravity forces on the wave, we eliminate the back-reaction of the aerodynamic drag of the dust particles on the gas, which is akin to considering a dust-gas mixture with a very low dust-to-gas mass ratio, $\mu$ (see equation~\ref{eq:gasmom}). The equations that describe the full self-gravitating dust fluid in this case are thus,
\begin{subequations}
\label{eq:appdustsysmu}
\begin{equation}
\frac{\text{d} \rho_d}{\text{d}t} = -\rho_g(\nabla \cdot \boldsymbol{v}),
\end{equation}
\begin{equation}
\frac{\text{d} \boldsymbol{v}}{\text{d}t} = - \nabla\Phi  - \frac{\Omega}{\tau_s}(\boldsymbol{v}-\overline{\boldsymbol{u}}),
\end{equation}
\begin{equation}
\nabla^2 \Phi = 4\pi G\rho_d,
\end{equation}
\end{subequations}
where $\overline{\boldsymbol{u}}$ is the background gas velocity, but going forward we will set this velocity to zero, placing the integration in the frame of the background gas fluid, leaving the drag term above proportional only the to dust fluid velocity w.r.t. to this background, stationary gas fluid.The other symbols represent the same quantities as Section~\ref{sec:nummeth}. In the frame of the shearing flow, given by $((3/2)\Omega x)\hat{\bs{y}}$, we have,
\begin{subequations}
\label{eq:appdustSS}
\begin{equation}
\frac{\text{d} \rho_d}{\text{d} t} - \frac{3}{2}\Omega x
\frac{\partial\rho_d}{\partial y} = -\rho_d\nabla\cdot\boldsymbol{v},
\end{equation}
\begin{equation}
\frac{\text{d}\boldsymbol{v}}{\text{d} t} -\frac{3}{2}\Omega x\frac{\partial\boldsymbol{v}}{\partial y} = 2\Omega v_y \hat{x}
-\frac{1}{2}\Omega v_x\hat{y} - \nabla\Phi - \frac{\Omega}{\tau_s}\boldsymbol{v},
\end{equation}
\begin{equation}
\nabla^2 \Phi = 4\pi G\rho_d.
\end{equation}
\end{subequations}
Now, following \citet{GLB65I}, we transform to sheared axes, which we denote with a ${}^\prime$,
\begin{subequations}
\begin{align}
x^\prime &= x  \\
y^\prime &= y + (3/2)\Omega xt \\
t^\prime &= t 
\end{align}
\end{subequations}
and the derivatives in terms of these axes are,
\begin{subequations}
\begin{align}
\frac{\partial}{\partial x} &= \frac{\partial}{\partial x^\prime} + (3/2)\Omega t^\prime\frac{\partial}{\partial y^\prime} \\
\frac{\partial}{\partial y} &= \frac{\partial}{\partial y^\prime} \\
\frac{d}{dt} &= \frac{d}{dt^\prime} + (3/2)\Omega x^\prime\frac{\partial}{\partial y\prime}.
\end{align}
\end{subequations}
The linear perturbations to the fluid properties are of the form,
\begin{equation}
\rho_d = \rho_{d0}[1 + \delta\rho_d],
\end{equation}
\begin{equation}
\boldsymbol{v} = \overline{\boldsymbol{v}} + \delta\boldsymbol{v},
\end{equation}
thus, we note that $\delta \rho_d$ is a dimensionless quantity. The functional form the perturbations is a plane wave in the sheared axes,
\begin{equation}
\delta f(x^\prime,y^\prime) = {\widetilde{f}\exp[i(k_xx^\prime + k_yy^\prime -\omega t)]}.
\end{equation}
Lastly, still following \citet{GLB65I}, we denote a dimensionless shear time parameter
\begin{equation}
\label{eq:GLBtau}
\tau \equiv (3/2)\Omega t^\prime - k_x/k_y,
\end{equation}
and we will track the temporal evolution of the wave according to this parameter $\tau$. Returning to the shearing-frame fluid equations from eq.~\ref{eq:appdustSS}, applying the linear, small-amplitude perturbations and discarding non-linear terms, we have the equations that describe the evolution of the amplitudes of the wave with the dimensionless time $\tau$:
\begin{subequations}
\begin{equation}
\frac{d\delta\rho_d}{d\tau} = -i \frac{2 k_y}{3\Omega}(\delta v_x \tau + \delta v_y),
\end{equation}
\begin{equation}
\frac{d\delta v_x}{d\tau} = \frac{4}{3} \delta v_y 
 + i\frac{2}{3\Omega}\tau\frac{4\pi G}{k_y(1+\tau^2)}\rho_{d,0}\delta\rho_d - \frac{2}{3\tau_s}\delta v_x,
\end{equation}
\begin{equation}
\frac{d\delta v_y}{d\tau} = -\frac{1}{3} \delta v_x + i\frac{2}{3\Omega}\frac{4\pi G}{k_y(1+\tau^2)}\rho_{d,0}\delta\rho_d - \frac{2}{3\tau_s}\delta v_y.
\end{equation}
\end{subequations}
We choose the following parameters for the numerical (\textsc{Athena}) and semi-analytic integrations: $\tau_s = \rho_{d,0} = k_y = G = 1.0$, and the initial conditions: $\tau_0 = -2$, $\delta v_x(\tau_0) = \delta v_y(\tau_0) = 0$, $\delta \rho_d(\tau_0) = 10^{-6}$. The domain in \textsc{Athena} is set-up with ($L_x$,$L_y$,$L_z$) = ($2\pi$,$2\pi$,$0.2$) and ($N_x$,$N_y$,$N_z$) = ($256$,$256$,$2$). 

The evolution of the amplitudes of the sheared wave is shown in Figure~\ref{fig:appshearwave}. The numerical and semi-analytic solutions agree strongly until $\tau \sim 4$, when the $\delta\rho_d/\rho_{d,0}$ amplitude approaches $0.1$, and the perturbation becomes non-linear. At this point the linearized equations no longer describe the non-linear behavior captured in the numerical integration. This confirms that our implementation of self-gravity for the dust particles follows the expected behavior.


\bsp	
\label{lastpage}
\end{document}